\documentclass[12pt]{article}
 \usepackage[latin1]{inputenc}
 \usepackage{amssymb}
 \usepackage{amsmath}
 \usepackage{amsfonts}
 \usepackage{slashed}
 \usepackage{graphicx}
 \usepackage{setspace}
 \usepackage{fancyhdr}
\usepackage{geometry}
\usepackage{plain}
 \usepackage{multirow}
 \usepackage{graphicx}
 \usepackage{anysize}
\usepackage{color}
 \usepackage{setspace}
 \usepackage[backgroundcolor=green,linecolor=green]{todonotes}
 \usepackage{array}
\usepackage[colorlinks=true]{hyperref}
 \usepackage[all]{hypcap}
 \numberwithin{equation}{section}
 \hypersetup{colorlinks, citecolor=violet, linkcolor=violet,  urlcolor=violet}
 \marginsize{1cm}{1cm}{2cm}{2cm}

\begin{document}

\newcommand{\ket}[2]{\langle #1,#2\rangle}
\newcommand{\bra}[2]{\left[ #1,#2\right]}
\newcommand{\Pgen}[3]{\langle#1\lvert#2\lvert#3]}

\thispagestyle{empty}

\begin{center}
\vspace{15mm}
\Large{\textbf{Parity Symmetry and Soft Limit for the Cachazo-Geyer Gravity Amplitude}} \\
\vspace{20mm}
\large\text{Brenda Penante, Sayeh Rajabi, Grigory Sizov}\\
\vspace{15mm}

\normalsize \textit{Perimeter Institute for Theoretical Physics, Waterloo, ON, N2L 2Y5, CA} \\
\vspace{2mm}
$\&$ \\
\vspace{2mm}
\normalsize\textit{Department of Physics and Astronomy $\&$ Guelph-Waterloo Physics Institute,} \\
\normalsize \textit{University of Waterloo, Waterloo, ON, N2L 3G1, CA}\\

\let\thefootnote\relax\footnotetext{bpenante, srajabi, gsizov@perimeterinstitute.ca}
\end{center}
\vspace{20mm}

\abstract
In this note, we prove that the recent proposal for the tree-level $n$-particle $\mathcal{N}=8$ supergravity amplitudes by Cachazo and Geyer \cite{Cachazo:2012da} satisfies parity symmetry and soft limit behavior expected for graviton scattering amplitudes.

\newpage

\section{Introduction}
\setcounter{page}{1}

In the last few years, there have been many developments toward a new formulation for scattering amplitudes in gauge theory and gravity.
For $ \mathcal{N}=4 $ super Yang-Mills, some examples are the BCFW recursion relations \cite{BCF:RecursionRelations,BCFW:ProofBCFW}, CSW expansion \cite{Cachazo:2004kj} and the Grassmannian formulations as integrals over the Grassmannians $ G(2,n) $ and $ G(k,n) $ \cite{ArkaniCachazo:DualityS-matrix2009,ArkaniCachazo:UnificationRes2009,Bullimore:2009cb,Bourjaily:2010kw}. Together they provide new computational tools as well as new conceptual insights, by making manifest some symmetries which are hidden in the Lagrangian formulation of the theory.

The Lagrangian approach suggests that even the tree-level gravity is indeed a much more complex problem than Yang-Mills, in which a naive perturbative approach becomes computationally unfeasible already for a very small number of particles. Surprisingly, a huge development in gravity amplitudes has been recently made by Hodges \cite{Hodges:2012ym} to reformulate tree-level MHV amplitudes in terms of a determinant.   

Amazingly, triggered by Hodges MHV formula, two distinct novel formulas for tree-level supergravity amplitudes of all $ R $-charge sectors were recently proposed by Cachazo-Geyer \cite{Cachazo:2012da} and Cachazo-Skinner \cite{Cachazo:2012kg}.

Both proposals, which are analogous to the Witten-Roiban-Spradlin-Volovich's (Witten-RSV) twistor string formulation of $\mathcal{N}=4$ SYM \cite{Witten:GaugeAsStringInTwistor2003,Roiban:2004yf}, can be understood as huge steps toward finding a twistor-string formulation of gravity. Hodges-like determinants are important ingredients in the two formulas, which make them so simple and elegant. 

On the one hand, the Cachazo-Geyer formula was derived from the supersymmetric version of the Kawai-Lewellen-Tye (KLT) relations \cite{Kawai:1985xq} which relate the maximally supersymmetric amplitudes in Yang-Mills and gravity. The Cachazo-Skinner formula, on the other hand, emerged from studying the BCFW relations for gravity in super-twistor space. 

In this short note, we consider the Cachazo-Geyer proposal and study two consistency checks, namely the parity invariance and soft limit behavior of the formula. Our proofs use the results given by RSV \cite{Roiban:2004yf} and Witten \cite{Witten:2004cp} for the parity and soft limit checks in $\mathcal{N}=4$ SYM, and present the validity of these properties in the gravity formula. These two checks are strong evidences that the proposal is the complete tree-level S-matrix of supergravity. Finally, we explore the possibility of using the known results in SYM to compute amplitudes in gravity from the proposed formula, and we show the explicit computations for MHV and $\overline{\text{MHV}}$ amplitudes.

The paper is organized as follows: In section \ref{secCG} we review the Cachazo-Geyer and Witten-RSV formulas. The proofs for the parity invariance and soft limit of the formula are presented in sections \ref{secparity} and \ref{secSL} respectively. Finally, in section \ref{secExamples} we calculate the MHV and $\overline{\text{MHV}}$ amplitudes from the formula.

\section{The Cachazo-Geyer and Witten-RSV Formulas}
\label{secCG}
Analogous to the Witten-RSV formulation, the Cachazo-Geyer formula for $n$ particle tree-level supergravity amplitudes in the $k^{\text{th}}$ sector is
\begin{align}
\begin{split}
\label{eq:amplitude1}
\mathcal{M}_{n,k}=&\frac{1}{\mathrm{Vol}(\text{GL}(2))}\int d^{2n}\sigma d^{2k}\rho \frac{H_n(\sigma)}{J_n(\sigma,\rho)}\prod\limits_{\alpha=1}^k\delta^2\left(\sum\limits_{a=1}^n C^V_{\alpha a}(\sigma)\tilde{\lambda}_a\right)\\
&\times\delta^{0|8}\left(\sum\limits_{a=1}^n C^V_{\alpha a}(\sigma)\tilde{\eta}_a\right)\prod\limits_{a=1}^n\delta^2\left(\sum\limits_{\alpha=1}^k\rho_\alpha C^V_{\alpha a}(\sigma)-\lambda_a\right),\\
\end{split}
\end{align}
whose ingredients we explain below:\\
\begin{itemize}
\item $ C^V(\sigma) $ is a $ k\times n $ matrix obtained from the world-sheet variables $ (\sigma_{1a},\sigma_{2a}),\; a=1\ldots n $, via the Veronese map
\begin{align}
\label{eq:VeroneseMap}
\begin{split}
V:G(2,n)&\rightarrow G(k,n)\\
\Sigma_{2\times n}\;&\mapsto C^V(\sigma)_{k\times n}
\end{split}.
\end{align}
Each column of $ \Sigma $ transforms as
\begin{equation}
\begin{pmatrix}\sigma_{1a} \\ \sigma_{2a}\end{pmatrix} \overset{V}\mapsto\begin{pmatrix}(\sigma_{1a})^{k-1} \\ (\sigma_{1a})^{k-2}\sigma_{2a}\\ \vdots  \\ (\sigma_{2a})^{k-1}\end{pmatrix},\qquad C^V_{\alpha a}(\sigma)=(\sigma_{1a})^{k-\alpha}(\sigma_{2a})^{\alpha-1}.
\label{eq:Veronese}
\end{equation}\\
\item In order to write $ H_n $ let us first define the Hodges-like $ n\times n $ matrix $ \Phi_n $ with elements
\begin{align}
\label{eq:Phi}
\begin{split}
(\Phi_n)_{ab}&=\frac{s_{ab}}{(a\,b)^2},\qquad\text{for } a\neq b\\
(\Phi_n)_{aa}&=-\sum\limits_{\substack{b=1\\b\neq a}}^n\frac{s_{ab}}{(a\,b)^2}\frac{(b\,l)(b\,r)}{(a\,l)(a\,r)}. 
\end{split}
\end{align}
We recall to the reader that $ (a\,b) $ are $2\times 2$ minors of $ \Sigma $, defined as 
\begin{equation}
(a\,b)=\sigma_{1a}\sigma_{2b}-\sigma_{1b}\sigma_{2a}.
\end{equation}
The matrix $ \Phi_n $ has rank $ n-3 $. A non-degenerate matrix can be obtained by deleting three rows and three columns of $ \Phi_n $. This matrix is denoted by $  \Phi_{n(def)}^{\hspace{1.6mm}(abc)} $. Finally, $ H_n $ is defined as
\begin{equation}
H_n=(-1)^{n+1}\frac{1}{(a\,b)(b\,c)(c\,a)}\times\frac{1}{(d\,e)(e\,f)(f\,d)}|\Phi_{n(def)}^{\hspace{1.6mm}(abc)}|.
\end{equation}
\item The last ingredient $ J_n $ is obtained in a similar fashion as $ H_n $. First define the $ 2n+2k $ vectors grouping the sets of variables $ \mathcal{V} $ and equations $ \mathcal{E} $:
\begin{align}
\begin{split}
\mathcal{V}&=\{\rho_{11},\rho_{12},\ldots, \rho_{k1},\rho_{k2},\sigma_{11},\sigma_{21},\ldots,\sigma_{1n},\sigma_{2n}\}\\
\mathcal{E}&=\{E_{11},E_{12},\ldots, E_{k1},E_{k2},F_{11},F_{21},\ldots, F_{1n},F_{2n}\} 
\end{split}
\end{align}
where
\begin{equation}
 E_{\alpha\underline{\dot{\alpha}}}=\sum\limits_{a=1}^n C^V_{\alpha a}(\sigma)\tilde{\lambda}_{a\underline{\dot{\alpha}}},\qquad F_{\underline\alpha a}=\sum\limits_{\alpha=1}^k\rho_{\alpha\underline{\alpha}} C^V_{\alpha a}(\sigma),
\end{equation}
here $ \underline{\alpha},\underline{\dot{\alpha}}=1,2 $ are holomorphic and anti-holomorphic spinor indices, respectively.
We then construct the matrix 
\begin{equation}
 (K_n)_{IJ}=\frac{\partial\mathcal{E}_J}{\partial\mathcal{V}_I},\qquad K_n=\begin{pmatrix}
\left(\frac{\partial E}{\partial \rho}\right)_{2k\times 2k} & \left(\frac{\partial F}{\partial \rho}\right)_{2k\times 2n}\\
\left(\frac{\partial E}{\partial \sigma}\right)_{2n\times 2k} & \left(\frac{\partial F}{\partial \sigma}\right)_{2n\times 2n}
\end{pmatrix}.
\end{equation}
Again, we know that since the system of equations contains momentum conservation, its rank is $ 2n+2k-4 $. Therefore, a non-degenerate matrix can be obtained by deleting four rows and four columns of $ K_n $. Choosing the rows to be the ones corresponding to $ \{\sigma_{1a},\sigma_{2a},\sigma_{1b},\sigma_{2b}\} $ and the columns to $  \{F_{1c},F_{2c},F_{1d},F_{2d}\} $, and denoting the remaining matrix by $ K_{n(cd)}^{\hspace{1.6mm}(ab)} $, $ J_n $ is finally given by
\begin{equation}
\label{eq:jacobian}
J_n=\frac{1}{(a\,b)^2[c\,d]^2}|K_{n(cd)}^{\hspace{1.6mm}(ab)}|.     
\end{equation}
\end{itemize}
The number of integration variables in \eqref{eq:amplitude1} after gauge fixing the GL$ (2) $ redundancy is $ 2n+2k-4 $, the same number of $ \delta $-functions under the integral after pulling out the momentum conserving $ \delta^4\left(\sum_{a=1}^np_a\right) $. Hence, the integral is completely localized. The resulting amplitude is therefore computed from the solutions $ (\sigma_{1a}^*,\sigma_{2a}^*,\rho^*_{\alpha\underline{\alpha}}) $ of the system of equations
\begin{align}
\label{eq:system}
\sum\limits_{\alpha=1}^k\rho_\alpha C^V_{\alpha a}(\sigma)=\lambda_a, \qquad\qquad\sum\limits_{a=1}^n C^V_{\alpha a}(\sigma)\tilde{\lambda}_a=0,
\end{align}
by evaluating the integrand with the corresponding Jacobian factor at each solution and summing over all contributions. The Cachazo-Geyer formula can be finally written as
\begin{align}
\label{eq:amplitude2}
\begin{split}
\mathcal{M}_{n,k}&=\delta^4\left(\sum\limits_{a=1}^{n}p_a\right)\sum\limits_{\substack{\mathrm{Solutions}\\(\rho^*,\sigma_1^*,\sigma_2^*)}}\left.\frac{H_n}{J^2_n}\prod\limits_{\alpha=1}^k\delta^{0|8}\left(\sum\limits_{a=1}^n C^V_{\alpha a}(\sigma)\tilde{\eta}_a\right)\right|_{(\rho^*,\sigma_1^*,\sigma_2^*)}\\
&=\delta^4\left(\sum\limits_{a=1}^{n}p_a\right) M_{n,k}.\\ \end{split}
\end{align}
For the proofs presented in this work, it turns out to be more convenient to write \eqref{eq:amplitude1} in terms of other set of variables $ (\rho,\sigma,\xi) $ where $ \sigma$ and $\xi $ are related to the world-sheet variables $ \sigma_1$ and $\sigma_2 $ as
\begin{equation}
\xi_a=(\sigma_{1a})^{k-1},\qquad\sigma_a=\frac{\sigma_{2a}}{\sigma_{1a}}.
\end{equation}
In these variables, \eqref{eq:amplitude1} reads
\begin{align}
\begin{split}
\label{eq:amplitudegravRSV}
\mathcal{M}_{n,k}=&\frac{1}{\mathrm{Vol}(\text{GL}(2))}\int \frac{d^n\xi d^{n}\sigma d^{2k}\rho}{(k-1)^{2n}} \prod\limits_{a=1}^n\xi_a^{\frac{k-3}{k-1}} \frac{H_n(\sigma,\xi)}{J'_n(\sigma,\xi,\rho)}\prod\limits_{\alpha=1}^k\delta^2\left(\sum\limits_{a=1}^n C^V_{\alpha a}(\xi,\sigma)\tilde{\lambda}_a\right)\\
&\times\delta^{0|8}\left(\sum\limits_{a=1}^n C^V_{\alpha a}(\xi,\sigma)\tilde{\eta}_a\right)\prod\limits_{a=1}^n\delta^2\left(\sum\limits_{\alpha=1}^k\rho_\alpha C^V_{\alpha a}(\xi,\sigma)-\lambda_a\right),\\
\end{split}
\end{align}
where $ C^V_{\alpha a}(\xi,\sigma)=\xi_a\sigma_a^{\alpha-1}  $ and $ J_n' $ is the Jacobian obtained by solving the $ \delta $-functions with respect to the variables $ (\rho,\sigma,\xi) $, related to $ J_n $ as
\begin{equation}
J_n'=\frac{J_n}{(k-1)^n}\prod\limits_{a=1}^n\xi^{\frac{3-k}{k-1}}_a.
\end{equation}
In the following, we will use known results in $ \mathcal{N}=4 $ SYM theory by Witten-RSV to verify that \eqref{eq:amplitude2} indeed obeys parity invariance and reproduces the correct soft factor \cite{Bern:1998sv,PhysRev.140.B516}.\\
\newline
In order to do so, it is instructive to review the Witten-RSV formulation of gauge theory amplitudes in terms of Witten's twistor string \cite{Witten:GaugeAsStringInTwistor2003}. The SYM $ n $-point partial amplitudes in the $ k^{\text{th}} $ sector are given by
\begin{align}
\label{eq:RSV}
\begin{split}
\mathcal{A}_{n,k}(1,\ldots,n)=&\frac{1}{\text{Vol(GL(2))}}\frac{1}{(k-1)^n}\int \frac{d^n\sigma d^n\xi d^{2k}\rho}{\prod\limits_{a=1}^n\xi_a(\sigma_a-\sigma_{a+1})}\prod\limits_{\alpha=1}^k\delta^2\left(\sum\limits_{a=1}^n C^V_{\alpha a}(\xi,\sigma)\tilde{\lambda}_a\right)\\
&\times\delta^{0|8}\left(\sum\limits_{a=1}^n C^V_{\alpha a}(\xi,\sigma)\tilde{\eta}_a\right)\prod\limits_{a=1}^n\delta^2\left(\sum\limits_{\alpha=1}^k\rho_\alpha C^V_{\alpha a}(\xi,\sigma)-\lambda_a\right).
\end{split}
\end{align} 

Once again, this integral is completely localized and the amplitude is given by
\begin{align}
\label{eq:ampgauge}
\begin{split}
\mathcal{A}_{n,k}(1,\dots, n)&=\frac{\delta^4\left(\sum\limits_{a=1}^{n}p_a\right)}{(k-1)^n}\sum\limits_{\substack{\mathrm{Solutions}\\(\rho^*, \sigma^*, \xi^*)}}\left.\frac{1}{\prod\limits_{a=1}^n\xi_a(\sigma_a-\sigma_{a+1})}\frac{1}{J'_n}\prod\limits_{\alpha=1}^k\delta^{0|4}\left(\sum\limits_{a=1}^n C^V_{\alpha a}(\xi,\sigma)\tilde{\eta}_a\right)\right|_{(\rho^*, \sigma^*, \xi^*)}\\
&=\delta^4\left(\sum\limits_{a=1}^{n}p_a\right)A_{n,k}(1,\dots, n),
\end{split}
\end{align}
with $J'$ being the Jacobian obtained by solving equations \eqref{eq:system} with respect to the variables $(\rho, \sigma, \xi)$.

\section{Parity Invariance}
\label{secparity}

In this section we check the parity symmetry of the Cachazo-Geyer formula. Given a scattering amplitude, the parity conjugated one is obtained by swapping $ k\leftrightarrow n-k $, $ \lambda\leftrightarrow \tilde{\lambda} $ and Fourier transforming the Grassmann SUSY parameters $ \tilde{\eta}\leftrightarrow \eta $. Being more explicit, this statement for gravity reads
\begin{equation}
\label{eq:parity}
\mathcal{M}_{n,k}(\lambda,\tilde{\lambda},\tilde{\eta})=\int d^{8n}\eta \exp\left({i\sum\limits_{\substack{a=1\\A=1,\ldots, 8}}^{n}\tilde{\eta}^A_a\eta^A_a}\right)\widetilde{\mathcal{M}}_{k,n-k}(\tilde{\lambda},\lambda,\eta).
\end{equation}
We start from the result stated by RSV that for each solution $(\rho^*, \sigma^*, \xi^*)$ of the $ (n,k) $ system of equations \eqref{eq:system} there corresponds a solution $(\tilde{\rho},\tilde{\sigma}, \tilde{\xi})$ of the $ (n,n-k) $ system of equations of the parity conjugated amplitude
\begin{equation}
\label{eq:paritysystem}
\sum\limits_{\beta=1}^{n-k}\tilde\rho_\beta \tilde C^V_{\beta a}(\tilde \sigma)=\tilde\lambda_a, \qquad\qquad\sum\limits_{a=1}^n \tilde C^V_{\beta a}(\tilde \sigma)\lambda_a=0,
\end{equation}
where $\tilde{C}^V(\tilde{\sigma})$ is an $(n-k)\times n$ matrix also obtained via the Veronese map \eqref{eq:Veronese} from some $2\times n $ matrix $ \tilde{\Sigma} $. The conjugated solutions are obtained by performing the change of variables
\begin{align}
\label{eq:paritysolutions}
\tilde{\sigma}_a=\sigma_a,\qquad\qquad\tilde{\xi}_a=\frac{1}{\xi_a\prod\limits_{b\neq a}^{n}(\sigma_a-\sigma_b)}.
\end{align}
From now on we denote $ \sigma_{ab}\equiv \sigma_a-\sigma_b $.
We want to show that each individual solution is parity invariant, that is
\begin{align}
\label{eq:amplitude3}
\begin{split}
\left.\frac{H_n}{J^2_n}\prod\limits_{\alpha=1}^k\delta^{0|8}\left(\sum\limits_{a=1}^n C^V_{\alpha a}(\sigma)\tilde{\eta}_a\right)\right|_{(\rho^*,\sigma_1^*,\sigma_2^*)}=\left.\int d^{8n}\eta e^{i\eta_a\tilde{\eta}^a}\frac{\tilde H_n}{\tilde{J}^2_n}\prod\limits_{\beta=1}^{n-k}\delta^{0|8}\left(\sum\limits_{a=1}^n \tilde C^V_{\beta a}\eta_a(\tilde{\sigma})\right)\right|_{(\tilde\sigma_1^*,\tilde\sigma_2^*,\tilde\rho^*)},
\end{split}
\end{align}
where $ (\tilde\sigma_1^*,\tilde\sigma_2^*,\tilde\rho^*) $ are the solutions of \eqref{eq:paritysystem}.\\
\newline
In SYM, the correspondence between individual solutions leads to the identity
\begin{equation}
\label{eq:identity1}
\left.\frac{\prod\limits_{\alpha=1}^{k}\delta^{0|4}\left(\sum\limits_{a=1}^n C^V_{\alpha a}(\sigma)\tilde{\eta}_a\right)}{(k-1)^{n}J'\prod\limits_{a=1}^n\xi_a\sigma_{a+1\,a}}\right|_{(\rho^*, \sigma^*, \xi^*)}=\left.\int d^{4n} \eta e^{i\eta_a\tilde{\eta}^a}\frac{\prod\limits_{\beta=1}^{n-k}\delta^{0|4}\left(\sum\limits_{a=1}^n \tilde C^V_{\beta a}(\tilde{\sigma})\eta_a\right)}{(n-k-1)^{n}\tilde J'\prod\limits_{a=1}^n \tilde\xi_a\tilde{\sigma}_{a+1\,a}}\right|_{(\tilde{\rho}^*, \tilde{\sigma}^*, \tilde{\xi}^*)}.
\end{equation}
We see immediately that the factor $ \prod_{a=1}^n \sigma_{a+1\,a}=\prod_{a=1}^n\tilde{\sigma}_{a+1\,a} $ cancels in both sides as a consequence of \eqref{eq:paritysolutions} and the equality between each solution.
This identity when written in terms of the variables $ (\rho,\sigma_1,\sigma_2) $ as in \eqref{eq:amplitude1} reads
\begin{align}
\label{eq:identity2}
\begin{split}
\left.\prod\limits_{\alpha=1}^k\delta^{0|4}\left(\sum\limits_{a=1}^n C^V_{\alpha a}(\sigma)\tilde{\eta}_a\right)\prod\limits_{a=1}^n\xi_a^{-\frac{2}{k-1}}J^{-1}\right|_{(\rho^*,\sigma^*_1,\sigma^*_2)}=\\
\left.\int d^{4n} \eta e^{i\eta_a\tilde{\eta}^a}\prod\limits_{\beta=1}^{n-k}\delta^{0|4}\left(\sum\limits_{a=1}^n \tilde C^V_{\beta a}(\tilde{\sigma})\eta_a\right)\prod\limits_{a=1}^n\tilde{\xi}_a^{-\frac{2}{n-k-1}}\tilde{ J}^{-1}\right|_{(\tilde{\rho}^*,\tilde{\sigma}^*_1,\tilde{\sigma}^*_2)}.
\end{split}
\end{align}
The idea is to split the fermionic $ \delta^{0|8} $ from \eqref{eq:amplitude3} into two copies of $ \delta^{0|4} $, and then apply \eqref{eq:identity2} twice.
Now it is left to work out what $ H_n $ is in terms of the coordinates $ (\xi,\sigma) $.
Each $ 2\times 2 $ minor is given by $ (i\,j)=(\xi_i\xi_j)^{\frac{1}{k-1}}\sigma_{ji} $, thus we can factorize the $ \xi $ dependence of the matrix $ \Phi_n(\xi,\sigma) $ and write instead $ \Phi_n'(\sigma) $:
\begin{align}
(\Phi_n)_{ab}&=(\xi_a\xi_b)^{-\frac{2}{k-1}}\frac{s_{ab}}{(\sigma_{ab})^2}=(\xi_a\xi_b)^{-\frac{2}{k-1}}(\Phi'_n)_{ab}(\sigma),\qquad\text{for } a\neq b\\
(\Phi_n)_{aa}&=-\xi_a^{-\frac{4}{k-1}}\sum\limits_{b\neq a}\frac{s_{ab}}{(\sigma_{ab})^2}\frac{\sigma_{lb}\sigma_{rb}}{\sigma_{la}\sigma_{ra}}=\xi_a^{-\frac{4}{k-1}}(\Phi'_n)_{aa}(\sigma).
\end{align}
With this, $ H_n $ is given by
\begin{align}
\label{eq:Hn}
\begin{split}
H_n=&(-1)^{n+1}\prod\limits_{i=1}^n\frac{\xi_i^{-\frac{4}{k-1}}}{(\xi_a\xi_b\dots\xi_f)^{\frac{2}{k-1}}}\frac{1}{(\xi_a\xi_b\ldots\xi_f)^{-\frac{2}{k-1}}(\sigma_{ba}\ldots\sigma_{df})}|\Phi_n'(\sigma)^{(abc)}_{(def)}|\\
=&(-1)^{n+1}\prod\limits_{i=1}^n\xi_i^{-\frac{4}{k-1}}\frac{|\Phi_n'(\sigma)^{(abc)}_{(def)}|}{(\sigma_{ba}\ldots\sigma_{df})}.
\end{split}
\end{align}
Splitting $ \tilde{\eta}^A,\;A=1,\ldots, 8 $, into two Grassmann parameters $ \tilde\eta^A_\text{L},\tilde\eta^A_\text{R},\;A=1,\ldots ,4 $, we can write
\begin{align}
\begin{split}
\frac{H_n}{J^2_n}\prod\limits_{\alpha=1}^k\delta^{0|8}\left.\left(\sum\limits_{a=1}^n C^V_{\alpha a}(\sigma)\tilde{\eta}_a\right) \right|_{(\rho^*,\sigma^*_1,\sigma^*_2)}=H_n\prod\limits_{\alpha=1}^k\left(\frac{\delta^{0|4}\left(\sum\limits_{a=1}^n C^V_{\alpha a}(\sigma)\tilde{\eta}^a_\text{L}\right)}{J_n}\right)\times\\
\left.\left(\frac{\delta^{0|4}\left(\sum\limits_{a=1}^n C^V_{\alpha a}(\sigma)\tilde{\eta}^a_\text{R}\right)}{J_n}\right)\right|_{(\rho^*,\sigma^*_1,\sigma^*_2)}.
\end{split}
\end{align}
Then, using \eqref{eq:identity2} twice and merging the two Grassmann $ \eta_{\text{L,R}}^A,\;A=1,\ldots,4 $, conjugated to $ \tilde{\eta}^A_{L,R} $, into an 8-component $ \eta^A $, we get
\begin{align}
\left.\frac{H_n}{J^2_n}\prod\limits_{\alpha=1}^k\delta^{0|8}\left(\sum\limits_{a=1}^n C^V_{\alpha a}(\sigma)\tilde{\eta}_a\right)\right|_{(\rho^*,\sigma^*_1,\sigma^*_2)}=\left.\int d^{8n} \eta e^{i\eta_a\tilde{\eta}^a}\frac{\tilde{H}_n}{\tilde{J}^2_n}\prod\limits_{\beta=1}^{n-k}\delta^{0|8}\left(\sum\limits_{a=1}^n \tilde C^V_{\beta a}(\tilde{\sigma})\eta_a\right)\right|_{(\tilde{\rho}^*,\tilde{\sigma}^*_1,\tilde{\sigma}^*_2)},
\end{align}
with 
\begin{equation}
 \tilde{H}_n=- \prod\limits_{a=1}^n\tilde{\xi}_a^{-\frac{2}{n-k-1}}\frac{|\Phi_n'(\sigma)^{(abc)}_{(def)}|}{(\sigma_{ba}\ldots\sigma_{df})}.
\end{equation}
We conclude that each solution is invariant under parity transformation. This implies that the whole amplitude satisfies \eqref{eq:parity}.

\section{Soft Graviton Limit}
\label{secSL}

In this section, we show that the Cachazo-Geyer formula reproduces the correct soft factor for gravity amplitudes. In order to do so, we first recall to the reader the soft limit for SYM amplitudes: if particle 1 has positive helicity and we take its momentum to zero, the amplitude factorizes in the following way:
\begin{align}
A_n\xrightarrow{p_1\rightarrow 0} \frac{\langle 2\,n\rangle}{\langle n\,1\rangle\langle 1\,2\rangle}A_{n-1}.
\label{YM_soft}
\end{align}
If particle 1 has negative helicity, we simply conjugate the soft factor.\\
\newline
We start from the RSV formula \eqref{eq:ampgauge} in terms of the variables $ (\rho,\sigma_1,\sigma_2) $:
\begin{align}
A_{n,k}=\sum\limits_{\substack{\mathrm{Solutions}\\(\rho^*,\sigma_1^*,\sigma_2^*)}}\left.\prod\limits_{a=1}^n\frac{\xi_a^{-\frac{2}{k-1}}}{\sigma_{a+1\,a}}\frac{1}{J_n}\prod\limits_{\alpha=1}^k\delta^{0|4}\left(\sum\limits_{a=1}^n C^V_{\alpha a}(\sigma)\tilde{\eta}_a\right)\right|_{(\rho^*,\sigma_1^*,\sigma_2^*)}.
\end{align}
Under the soft limit $ (\lambda_1,\tilde{\lambda}_1)\rightarrow (0,0)$ and $\tilde{\eta}_1=0$, $J_n$ factorizes into 
\begin{equation}
\label{eq:Jfactor}
J_n=J_{n-1}D,
\end{equation}
with $ D $ being a $ 2\times 2 $ matrix that carries all dependence on the soft particle. To see this, let us recall from \eqref{eq:jacobian} the definition $ J_n=\frac{1}{(a\,b)^2[c\,d]^2}|K_{n(cd)}^{\hspace{1.6mm}(ab)}| $ and look at the matrix $ K_n $ under such limit. If we arrange the columns and rows in order to put the dependence on the soft particle on the last two ones, then
\begin{align}
K_n=\begin{pmatrix}
K_{n-1} & \vdots\\
A & D
\end{pmatrix},
\end{align}
where $ A $ is a $ 2\times 2(k+n-1) $ matrix in which all non-zero entries are of the kind  $ \frac{\partial E_{\alpha\underline{\dot{\alpha}}}}{\partial(\sigma_{11},\sigma_{21})}\propto\tilde{\lambda}^{\underline{\dot{\alpha}}}_1 $. Therefore, in the soft limit all its entries become zero and
\begin{equation}
 \det K_n =\det K_{n-1} \det D. 
\end{equation}
Choosing the deleted rows and columns $ \{a,b,c,d\}\neq 1 $, the factorization of $ |K_{n(cd)}^{\hspace{1.6mm}(ab)}| $ translates into the factorization \eqref{eq:Jfactor} of $ J_n $.\\
\newline
We extract the factors depending on the soft particle, and use the soft limit of the RSV formula \eqref{eq:ampgauge}
\begin{align}
\label{ymsoft}
A_{n,k}=\left.\sum_{\substack{\mathrm{Solutions}\\(\rho^*,\sigma^*,\xi^*)}}\xi^{-\frac{2}{k-1}}_1\frac{\sigma_{2n}}{\sigma_{1n}\sigma_{21}}\prod\limits^n_{a=2}\frac{1}{\xi_a(\sigma_{a+1a})}\frac{\prod\limits_{\alpha=1}^k\delta^{0|4}\left(\sum\limits_{a=1}^n C_{\alpha a}(\sigma,\xi)\tilde{\eta}_a\right)}{J_{n-1}}\frac{1}{D}\right|_{(\rho^*,\sigma^*,\xi^*)}, 
\end{align}
where we multiplied and divided by $ \sigma_{2n} $\footnote{Not to be confused with the elements of the $ 2\times n $ matrix $ \Sigma $ of \eqref{eq:VeroneseMap}.} in order to obtain the correct measure for $ A_{n-1} $.\\
\newline
Comparing \eqref{ymsoft} with (\ref{YM_soft}), we can find the factor $D$ at each solution
\begin{equation}
\label{D}
\frac{1}{D}=\left.\frac{\langle 2\,n\rangle}{\langle 1\,2\rangle\langle n\,1\rangle}\times\frac{\xi_1^{\frac{2}{k-1}}\sigma_{n1}\sigma_{12}}{\sigma_{n2}}\right|_{(\rho^*,\sigma^*,\xi^*)}=\left.\frac{\langle 2\,n\rangle}{(2\,n)}\frac{(1\,2)}{\langle 1\,2\rangle}\frac{(n\,1)}{\langle n\,1\rangle}\right|_{(\rho^*,\sigma_1^*,\sigma_2^*)},
\end{equation}
which will be useful below in the calculation of the soft limit. It is crucial to notice that $D$ does not know anything about the ordering of the particles, so in \eqref{D} we can replace $2$ and $n$ by any other two particles.
We will now make use of these facts to calculate the soft limit in gravity. Recall the Cachazo-Geyer formula \eqref{eq:amplitude2}
\begin{align}
M_{n,k}=\sum\limits_{\substack{\mathrm{Solutions}\\(\rho^*,\sigma_1^*,\sigma_2^*)}}\left.\frac{H_n}{J^2_n}\prod\limits_{\alpha=1}^k\delta^{0|8}\left(\sum\limits_{a=1}^n C^V_{\alpha a}(\sigma)\tilde{\eta}_a\right)\right|_{(\rho^*,\sigma_1^*,\sigma_2^*)}.
\end{align}
In the soft limit $H_n$ also factorizes as
\begin{align}
\label{eq:factorH2}
H_n=H_{n-1}\sum\limits_{i=2}^n\frac{s_{1i}(i\,l)(i\,r)}{(a\,i)^2(a\,l)(a\,r)}.
\end{align}
To see this, let us recall from \eqref{eq:Hn} 
\[ H_n=(-1)^{n+1}\prod\limits_{i=1}^n\xi_i^{-\frac{4}{k-1}}\frac{|\Phi'_n(\sigma)^{(abc)}_{(def)}|}{(\sigma_{ba}\ldots\sigma_{df})}.\]
The determinant 
$ |\Phi'_n(\sigma)^{(abc)}_{(def)}| $ in the limit $ (\lambda_1,\tilde{\lambda}_1)\rightarrow (0,0) $ can be approximated as
\begin{align}
\label{eq:factorH}\begin{split}
|\Phi'_n(\sigma)^{(abc)}_{(def)}|=
  \left|\begin{array}{c|ccc}
    -\sum\limits_{i=2}^n\frac{s_{1i}}{(\sigma_{1\,i})^2}\frac{\sigma_{l\,i}\sigma_{r\,i}}{\sigma_{l\,1}\sigma_{r\,1}} & \frac{s_{12}}{(12)^2} & \cdots & \frac{s_{1n}}{(1\,n)^2} \\ \hline
    \frac{s_{21}}{(21)^2}  & \multicolumn{3}{c}{\multirow{3}{*}{\raisebox{-7mm}{\scalebox{1}{$\Phi'_{n-1}(\sigma)^{(abc)}_{(def)}$}}}} \\
    \raisebox{2pt}{\vdots} & & &\\
    \frac{s_{n1}}{(n\,1)^2} & & & 
  \end{array}\right|
\approx-\sum\limits_{i=2}^n\frac{s_{1i}}{(\sigma_{1\,i})^2}\frac{\sigma_{l\,i}\sigma_{r\,i}}{\sigma_{l\,1}\sigma_{r\,1}}|\Phi'_{n-1}(\sigma)^{(abc)}_{(def)}|
  \end{split}
\end{align}
by expanding in the first row and keeping only the first order in $ s_{1i} $, which is small in this limit.
Here we assumed that the set of three rows and three columns that are deleted from $ \Phi_n $ do not contain the first row or first column.
The $ \sigma_{ij} $ of \eqref{eq:factorH} combine with the $ \xi $ dependence of \eqref{eq:Hn} to give \eqref{eq:factorH2}. Thus, in the soft limit, the gravity amplitude takes the form
\begin{align}
M_{n,k}=\left.\sum\limits_{\substack{\mathrm{Solutions}\\(\rho^*, \sigma_1^*, \sigma^*_2)}}\sum_{c=2}^{n}\frac{s_{1c}(c\,l)(c\,r)}{(1\,c)^2(1\,l)(1\,r)}\frac{H_{n-1}}{J_{n-1}^2}\frac{1}{D^2}\right|_{(\rho^*,\sigma_1^*,\sigma_2^*)},
\end{align}
where the factor $H_{n-1}/J_{n-1}^2$ gives the lower-point amplitude $M_{n-1,k}$.

Now we can use the expression (\ref{D}) for $D$ with convenient replacements for the labels $2$ and $n$. There are two copies of $D$ in the formula, so choosing for the first one $\{2,n\}\rightarrow\{c,l\}$ and for the second one $\{2,n\}\rightarrow \{c,r\}$, we obtain
\begin{align}
\begin{split}
M_{n,k}&=\sum\limits_{\substack{\mathrm{Solutions}\\(\rho^*,\sigma_1^*,\sigma_2^*)}}\left.\sum_{c=2}^{n}\frac{s_{1c}(c\,l)(c\,r)}{(1\,c)^2(1\,l)(1\,r)}\frac{H_{n-1}}{J_{n-1}^2}\left(\frac{\langle c\,l\rangle}{(c\,l)}\frac{(1\,c)}{\langle 1\,c\rangle}\frac{(l\,1)}{\langle l\,1\rangle}\right)\left(\frac{\langle c\,r\rangle}{(c\,r)}\frac{(1\,c)}{\langle 1\,c\rangle}\frac{(r\,1)}{\langle r\,1\rangle}\right)      \right|_{(\rho^*,\sigma_1^*,\sigma_2^*)}\\
&=\sum\limits_{\substack{\mathrm{Solutions}\\(\rho^*,\sigma_1^*,\sigma_2^*)}}\sum_{c=2}^{n}\left.\frac{[1c]\langle cl\rangle \langle cr\rangle}{\langle 1c\rangle\langle 1l\rangle \langle 1r\rangle}\times M_{n-1,k}( 2,\ldots, n)\right|_{(\rho^*,\sigma_1^*,\sigma_2^*)}.
\end{split}
\end{align}
Since for \emph{every} solution we obtain a factor that depends only on the external data, the full amplitude obtains the same factor. In other words, we have obtained the well-known expression for the gravitational soft limit
\begin{align}
\lim_{p_{1^+}\to 0}\mathcal{M}_{n,k}(1^+, 2,\ldots, n)&=\sum_{c=2}^{n}\frac{[1c]\langle cl\rangle \langle cr\rangle}{\langle 1c\rangle\langle 1l\rangle \langle 1r\rangle}\times \mathcal{M}_{n-1,k}( 2,\ldots, n).
\end{align}

\section{Calculating Gravity Amplitudes from SYM Results}
\label{secExamples}
It can be noticed that in both proofs in the previous sections we avoided calculating the clumsiest part of the formula \eqref{eq:amplitude1} ~--- the Jacobian $J_n$. This was achieved by making use of the knowledge of the corresponding SYM result. One can ask if it is possible to use the same trick also for computing amplitudes. It is trivially true for MHV, because the Jacobian is $1$ in this case. In this section, we will show that it also works for $\overline{\text{MHV}}$ whose Jacobian is not trivial. Obviously, $\overline{\text{MHV}}$ amplitudes can be obtained from the Hodges' MHV formula \cite{Hodges:2012ym} by parity conjugation. However, in order to illustrate using SYM results for calculating gravity amplitudes, we will show explicitly how the formula (\ref{eq:amplitude1}) reproduces the $\overline{\text{MHV}}$ amplitudes.\\
\newline
According to Hodges, a reduced\footnote{Following Hodges, we call reduced amplitude an amplitude with stripped momentum conserving $\delta$-function and Grassmannian $\delta$-functions.} tree-level MHV amplitude in  $ \mathcal{N}=8 $ SUGRA is given by 
\begin{align}
\bar{M}(1,2,\ldots,n)=(-1)^{n+1}\sigma(ijk,rst)\frac{|\Phi^H|^{rst}_{ijk}}{\langle ij\rangle \langle jk\rangle \langle ki\rangle\langle rs\rangle\langle st\rangle\langle tr\rangle},
\label{eq:HodgesFormula}
\end{align}
where \[\sigma(ijk,rst)=\text{sgn}((ijk12\dots\slashed i \slashed j \slashed k\ldots n)\rightarrow(rst12\ldots\slashed r \slashed s \slashed t\ldots n),\]
and $|\Phi^H|^{rst}_{ijk}$ is the $(n-3)\times(n-3)$ minor of the matrix 
\begin{align}
\label{eq:Hodges}
(\Phi^H)^i_j=\frac{[ij]}{\langle ij\rangle}, \  i\ne j,\qquad\qquad(\Phi^H)^i_i=-\sum\limits_{j\ne i}\frac{[ij]\langle jx\rangle\langle jy\rangle}{\langle ij\rangle \langle ix\rangle \langle iy\rangle},
\end{align}
obtained by deleting the columns $r,s,t$ and rows $i,j,k$. Here $x$ and $y$ are two arbitrary spinors.

First let us recall that each of the integral formulas for SYM (\ref{eq:RSV}) and gravity (\ref{eq:amplitude1})  can be written as a sum over solutions of the $ \delta $-functions in the integrand (\ref{eq:amplitude2},\ref{eq:ampgauge}). In the MHV case the integral receives contribution only from one solution, which by using the GL(2) ``gauge freedom'' can be written as
\begin{align}
\begin{pmatrix}
\sigma_{1a} &\ldots & \sigma_{1n} \\
\sigma_{2a} &\ldots & \sigma_{2n} \\
\end{pmatrix}=\begin{pmatrix}
\lambda_1^a &\ldots & \lambda_1^n \\
\lambda_2^a &\ldots & \lambda_2^n \\
\end{pmatrix}
,\qquad  \rho^1=\begin{pmatrix}
1\\
0
\end{pmatrix},\qquad \rho^2=\begin{pmatrix}
0\\
1
\end{pmatrix}
,
\label{eq:MHVsol}
\end{align}
or equivalently, in terms of $(\rho,\sigma,\xi)$
\begin{align}
\begin{cases}
\xi^a&=(\lambda^a_1)^{k-1}\\
\sigma^a&=\lambda^a_2/\lambda^a_1\\
\rho_\alpha^\beta&=\delta_\alpha^\beta
\end{cases}.
\end{align}
Therefore, on this solution the minors $(\sigma_a\sigma_b)$ become inner products $\langle a\,b\rangle$ and the matrix $\Phi_n$ of (\ref{eq:Phi})
reduces to the Hodges' matrix $\Phi^H$ (\ref{eq:Hodges}).

 On the MHV solution (\ref{eq:MHVsol}), $J_n=1$. Indeed, $J_n$ appears as a determinant of resolving the $\delta$-functions in (\ref{eq:amplitude1}), which in the MHV case takes the form
\begin{align}
\int d^{2n}\sigma d^2\rho_1 d^2\rho_2\, \delta^2\left(\sum\limits_{a=1}^n\sigma_{1a}\tilde\lambda_a\right)\delta^2\left(\sum\limits_{a=1}^n\sigma_{2a}\tilde\lambda_a\right)
\prod\limits_{a=1}^n\delta^2\left(\rho_1\sigma_{1a}+\rho_2\sigma_{2a}-\lambda^a\right).
\label{eq:deltasMHV}
\end{align}
On the solution, $\lambda^a=\sigma^a$, so the first two $\delta$-functions combine to the momentum conserving $\delta^4\left(\sum\limits_{a=1}^n\lambda^a_{\underline\alpha}\tilde\lambda^a_{\underline{\dot\alpha}}\right)$.
The last $2n$ $\delta$-functions in (\ref{eq:deltasMHV}) integrated over $\sigma$ can be written in a form 
\begin{align}
(\det R)^{-n}\int d^{2n}\sigma \prod\limits_{a=1}^n \delta^2(\sigma^a-R^{-1}\lambda^a)=(\det R)^{-n},\qquad R=\begin{pmatrix}
\rho^1_{\underline 1} & \rho^2_{\underline 1} \\
\rho^1_{\underline 2} & \rho^2_{\underline 2}
\end{pmatrix}.
\end{align}
But on the MHV solution, $R$ is equal to the identity matrix, so no factor is produced in (\ref{eq:deltasMHV}) and $J_n=1$.

 Substituting $J_n=1$ into the Cachazo-Geyer formula (\ref{eq:amplitude2}) and pulling out the momentum conserving and Grassmannian $\delta$-functions, we see that the reduced amplitude is equal to $H_n$. As (\ref{eq:Phi}) reduces to the Hodges' matrix (\ref{eq:Hodges}),
 we conclude that the Cachazo-Geyer formula (\ref{eq:amplitude1}) reproduces the Hodges' formula (\ref{eq:HodgesFormula}) in the MHV case.\\
\newline
Now we consider $\overline{\text{MHV}}$ amplitudes.
 First let us understand how (\ref{eq:RSV}) reproduces the well-known Parke-Taylor formula for MHV amplitudes in SYM
\begin{align}
A_{\text{MHV}}=\frac{\prod\limits_{\alpha=1}^2 \delta^{0|4}\left(\sum\limits_{a=1}^nC_{\alpha a}(\sigma)\tilde\eta_a\right)}{J_n\prod\limits_{a=1}^n (\sigma_a \sigma_{a+1})}
=\frac{\delta^{0|4}\left(\sum\limits_{a=1}^n\lambda^1_a\tilde\eta^a\right)\delta^{0|4}\left(\sum\limits_{a=1}^n\lambda^2_a\tilde\eta^a\right)}{\langle 12\rangle\ldots\langle n1\rangle }.
\end{align}
The corresponding formula for $\overline{\text{MHV}}$ can be obtained from MHV by parity conjugation. The only solution contributing to the amplitude is
\begin{align}
A_{\overline{\text{MHV}}}=\frac{\prod\limits_{\beta=1}^{n-2} \delta^{0|4}\left(\sum\limits_{a=1}^n\tilde C_{\beta a}(\tilde \sigma)\tilde\eta_a\right)}{\tilde J_n\prod\limits_{a=1}^n \left(\tilde\sigma_a \tilde\sigma_{a+1}\right)}=\int d^{4n}\eta\, 
e^{i\eta_a\tilde \eta^a}
\frac{\delta^{0|4}\left(\sum\limits_{a=1}^n\tilde\lambda^1_a\eta^a\right)\delta^{0|4}\left(\sum\limits_{a=1}^n\tilde\lambda^2_a\eta^a\right)}{[12]\ldots[n1]}.
\label{YM_MHVbar}
\end{align}
In order to make it clear that this is the conjugate of an MHV amplitude, we wrote the LHS in terms of the $(n-2)\times n$ matrix $\tilde{C}_{\beta a}=\tilde{\xi}_a\tilde{\sigma}_a^{\beta-1}$ where $\begin{pmatrix}
\tilde\sigma_1 \\
\tilde\sigma_2
\end{pmatrix}=\tilde\xi^{\frac{1}{n-3}}\begin{pmatrix}
1 \\
\tilde\sigma
\end{pmatrix} $
is a solution of the parity conjugated system \eqref{eq:paritysystem}.

Through the transformation (\ref{eq:paritysolutions}), $ \tilde{\sigma} $ and $ \tilde{\xi} $ are related to $\xi$ and $\sigma$, which in turn can be expressed through the antiholomorphic part of kinematical data
\begin{align}
\begin{split}
\xi^a&=(\tilde\lambda^a_1)^{n-3},\\
\sigma^a&=\frac{\tilde\lambda^a_2}{\tilde\lambda^a_1}.
\end{split}
\end{align}
We will use our knowledge of the $\overline{\text{MHV}}$ amplitudes in SYM to calculate them in gravity.
As in the previous sections, we represent the formula (\ref{eq:amplitude1}) for gravity amplitudes as a sum over solutions. Similar to the MHV case, there is only one solution
\begin{align}
M_{n,n-2}=\left.\frac{H_n}{J^2_n}\prod\limits_{\beta=1}^{n-2}\delta^{0|8}\left(\sum\limits_{a=1}^n C_{\beta a}(\sigma)\tilde\eta^a\right)\right|_{(\rho^*,\sigma_1^*,\sigma_2^*)}.
\end{align}
We split the $8$-component fermionic $\delta$-function into two $4$-component ones, group each of them with one copy of $1/J_n$ and use the SYM result (\ref{YM_MHVbar}). The two fermionic integrals over $d^{4n}\eta$ can be merged into one over $d^{8n}\eta$
\begin{align}
M_{n,n-2}= H_n\left(\frac{(\tilde\sigma_1\tilde\sigma_2)\ldots(\tilde\sigma_n\tilde{\sigma_1})}{[12]\ldots[n1]}\right)^2
\int d^{8n}\eta e^{i\eta_a\tilde \eta^a} \prod\limits_{\underline{\dot{\alpha}}=1}^2\delta^{0|8}\left(\sum\limits_{a=1}^n\tilde{\lambda}^a_{\underline{\dot{\alpha}}}\eta_a\right).
\label{Mn}
\end{align}
 Under the transformation (\ref{eq:paritysolutions}), $\sigma$ does not change while $\xi$ does, so it makes sense to extract the $\xi$-dependence from $H_n$, as in (\ref{eq:Hn}):
  \begin{align}
H_n=\prod\limits_{a=1}^n \tilde\xi^{-\frac{4}{n-k-1}}_a H'_n,
\label{eq:Hstripped}
 \end{align}
  where $H'_n$ depends only on $\sigma$ and thus does not change under the 
 transformation (\ref{eq:paritysolutions}). We can also extract the $\xi$-dependent factors from the minors

\begin{align}
\frac{(\tilde\sigma_1\tilde\sigma_2)\ldots(\tilde\sigma_n\tilde{\sigma_1})}{[12]\ldots [n1]}=\frac{\prod\limits_{a=1}^n \tilde\xi_a^{\frac{2}{n-k-1}}\tilde\sigma_{a+1\,1}}{\prod\limits_{a=1}^n \xi_a^{\frac{2}{k-1}}\sigma_{i+1\,1}}=
\prod\limits_{a=1}^n \frac{\tilde{\xi}_a^{\frac{4}{n-k-1}}}{\xi_a^{\frac{4}{k-1}} }.
\label{eq:minorsratio}
\end{align}
 Here we used the fact that 
 \begin{align}
 (\xi_a \xi_b)^{\frac{1}{k-1}}(\sigma_a-\sigma_b)=[a\,b].
 \label{eq:fact}
\end{align}
Substituting (\ref{eq:Hstripped}) and (\ref{eq:minorsratio}) into (\ref{Mn}), we see that the $\tilde{\xi}$'s cancel
\begin{align}
M_{n,n-2}=H'_n\prod\limits_{a=1}^n\xi_a^{-\frac{4}{k-1}}\int d^{8n}\eta e^{i\eta_a\tilde \eta^a} \prod\limits_{\underline{\dot{\alpha}}=1}^2\delta^{0|8}\left(\sum\limits_{a=1}^n\tilde\lambda^a_{\underline{\dot\alpha}}\eta_a\right).
\end{align}
 The $\xi$ factor can be now absorbed into $H'_n$ by defining \[H^{\text{conj}}_n=H'_n\prod\limits_{a=1}^n \xi^{-\frac{4}{k-1}}_a.\]
 Then, $H^{\text{conj}}_n$ can be calculated as a determinant of the matrix $\Phi_{\text{conj}}$
\begin{align}
\Phi^{\text{conj}}_{ij}=\frac{\langle i\,j\rangle}{[i\,j]} , i\ne j \qquad \qquad \Phi_{ii}^{\text{conj}}=-\sum\limits_j \frac{\langle i\,j\rangle[j\,l][j\,r]}{[i\,j][i\,l][i\,r]},
\end{align}
with three rows and three columns eliminated.
Here we used again (\ref{eq:fact}).\\
\newline
Finally, we obtain the following formula for $\overline{\text{MHV}}$ gravity amplitudes:
\begin{align}
 M_{n,n-2}=H^{\text{conj}}_n\int d^{8n}\eta e^{i\eta_a \tilde\eta^a}\delta^{0|4}\left(\sum\limits_{a=1}^n\tilde\lambda^1_a\eta_a\right)\delta^{0|4}\left(\sum\limits_{a=1}^n\tilde\lambda^2_a\eta_a\right).
\end{align}
Taking the fermionic integrations explicitly, we can represent the answer in the final form
\begin{align}
 M_{n,n-2}=H^{\text{conj}}_n\delta^{0|8}\left(\sum_{a\ne b}^n[a\,b]\prod^n\limits_{c\ne a,b}\tilde\eta_c\right),
\end{align}
which one can check that is the parity conjugate of Hodges' formula for MHV amplitudes \cite{Hodges:2012ym}.

We can conclude that in the MHV and $\overline{\text{MHV}}$ cases there is no need to explicitly calculate the Jacobian $J_n$ in the Cachazo-Geyer formula (\ref{eq:amplitude1}), because in the gravity $\overline{\text{MHV}}$ amplitude, $ J_n $ can be extracted from the SYM counterpart. This is a surprising fact, since unlike the MHV case, in which $J_n=1$, in the $\overline{\text{MHV}}$ case $ J_n $ is nontrivial, nevertheless this trick allows to avoid computing it explicitly. This simplification hints that there may be a possibility that the calculation of amplitudes with arbitrary $k$ does not require the computation of $ J_n $ explicitly, taking instead advantage of the corresponding SYM result. Therefore, this is a path for further simplifications of the formula \eqref{eq:amplitude1}.

\section{Conclusion}
In this paper, we have proved that the recently proposed Cachazo-Geyer formula (\ref{eq:amplitude1}) for all tree-level amplitudes in ${\cal N}=8$ SUGRA satisfies parity symmetry and behaves correctly in a soft-graviton limit. These properties provide evidence for the validity of the formula (\ref{eq:amplitude1}) in all $k$-sectors. Indeed, a $k$-preserving soft limit produces a lower-point amplitude with the same $k$. So, iteratively performing the $k$-preserving soft limit, each amplitude can be reduced to $\overline{\text{MHV}}$, which in turn can be related to MHV by parity conjugation. Thus, the consistency checks which we performed support the validity of the Cachazo-Geyer proposal.
\section*{Acknowledgments}
We are grateful to Freddy Cachazo for suggesting this problem, encouragement and many fruitful discussions in all stages of this work. It is also a pleasure to thank David Skinner for very useful conversations. Research at Perimeter Institute is supported by the Government of Canada through Industry Canada and by the Province of Ontario through the Ministry of Research \& Innovation.~S.R. is supported in part by the NSERC of Canada and MEDT of Ontario.

\bibliographystyle{utphys}
\bibliography{gravity}

\end{document}